\documentclass[aps,prl,twocolumn,superscriptaddress,showpacs]{revtex4-1}

\usepackage{amsmath,amssymb,bm}
\usepackage{graphicx}

\begin{document}


\title{Measurement of the noise spectrum using a multiple-pulse sequence}



\author{Tatsuro Yuge}
\email[]{yuge@m.tohoku.ac.jp}
\affiliation{
IIAIR, Tohoku University, Sendai 980-8578, Japan}

\author{Susumu Sasaki}
\affiliation{Department of Materials Science and Technology, Niigata University, 
Niigata 950-2181, Japan}

\author{Yoshiro Hirayama}
\affiliation{Department of Physics, Tohoku University, Sendai 980-8578, Japan}
\affiliation{ERATO Nuclear Spin Electronics Project, Sendai 980-8578, Japan}

\date{\today}

\begin{abstract}
A method is proposed for obtaining the spectrum for noise 
that causes the phase decoherence of a qubit
directly from experimentally available data. 
The method is based on a simple relationship between the spectrum 
and the coherence time of the qubit in the presence of a $\pi$-pulse sequence. 
The relationship is found to hold for every system of a qubit 
interacting with the classical-noise, bosonic, and spin baths.
\end{abstract}



\maketitle

{\it Introduction}.---Implementation of quantum information processing (QIP) 
requires maintaining the quantum coherence of a system during its operation.
In reality, however, systems are not completely isolated from the environment.
Interactions with the environment are system noise and cause decoherence.
One of the challenges in QIP implementation is 
to control the system appropriately in a noisy environment 
and/or to reduce the noise level by weakening the system-environment coupling.
Strategies for these approaches have been developed, 
including quantum error correction \cite{Shor, Steane, Preskill}, 
quantum estimation \cite{Helstrom, Watanabe_etal}, 
and dynamical decoupling (DD) 
\cite{ViolaLloyd, Facchi_etal, Uhrig, Uhrig2, CywinskiDasSarma_etal, YangLiu, PasiniUhrig, UhrigLidar, Yang_etal, Biercuk_etal, Lange_etal, BylanderTsai_etal}.  

In DD, the application of multiple $\pi$-pulses cancels out the noise 
and effectively suppresses the system-environment coupling \cite{ViolaLloyd}.
This basic idea of DD comes from concepts in pulsed 
nuclear magnetic resonance (NMR) \cite{Slichter}. 
Experiments have demonstrated that decoherence can be suppressed 
by using methods common to pulsed NMR, 
such as alternating-phase-Carr-Purcell (APCP) 
and Carr-Purcell-Meiboom-Gill (CPMG) methods 
\cite{WatanabeSasaki, DementyevBarrett_etal}.
Subsequent theoretical \cite{Facchi_etal, Uhrig, Uhrig2, CywinskiDasSarma_etal, YangLiu, PasiniUhrig, UhrigLidar, Yang_etal} 
and experimental \cite{Biercuk_etal, Lange_etal, BylanderTsai_etal} 
studies on DD in various kinds of spin- and charge-related qubit systems 
other than NMR compared several DD methods and suggested 
that optimization of DD requires knowledge of the noise properties.

For any strategy (not only DD), knowledge of the noise properties is helpful 
because it can be used to improve the strategy.
It is therefore important to identify the noise properties of the environment. 
For longitudinal decoherence (energy relaxation) noise, 
the noise spectrum can be obtained from 
the longitudinal relaxation time (called $T_1$ in NMR) \cite{BBP,Moriya,Slichter}, 
and a qubit (two-level system) has been proposed 
as a noise spectrum analyzer \cite{Schoelkopf_etal}.
In contrast, a method for measuring transverse decoherence (pure dephasing) noise 
has not been established 
although several methods (also using a qubit) have recently been proposed 
\cite{CywinskiDasSarma_etal, deSousa, Almog_etal}.
In Ref.~\cite{ deSousa}, it was pointed out that the relationship 
between noise spectrum and coherence can be used for estimating the spectrum.
In Ref.~\cite{CywinskiDasSarma_etal}, a method was described for obtaining 
the moments of the spectrum by using the Uhrig pulse sequence \cite{Uhrig}.
In Ref.~\cite{Almog_etal}, a method was described for obtaining the spectrum 
at the Rabi frequency by using a nearly-continuous and on-resonant control field.

In this Letter we propose a method for measuring the dephasing noise spectrum 
in which a simple sequence of equidistant $\pi$-pulses 
(such as an APCP or CPMG sequence) is used.
The spectrum at frequency $\pi/2\tau$, where $2\tau$ is the interval between pulses, 
is evaluated directly from experimentally obtained values of the coherence times  
by using a relationship between the noise spectrum and 
the coherence time for a sequence of a sufficiently large number of pulses. 
We show that this relationship holds for the classical-noise, spin-boson, 
and spin-spin bath models.

{\it Model and noise spectrum}.---We use a model of a single qubit (spin $S=1/2$) 
interacting with the environmental degrees of freedom (bath). 
The Hamiltonian of the total system is given by 
\begin{align}
\hat{H} = 
\frac{\hbar}{2} \bigl( \Omega + \hat{\xi} \bigr) \hat{\sigma}_z 
+ \hat{H}_{\rm bath},
\label{hamiltonian}
\end{align}
where $\hat{\xi}$ and $\hat{H}_{\rm bath}$ are the bath operators.
The Pauli matrices of the qubit are denoted by 
$\hat{\sigma}_x$, $\hat{\sigma}_y$, and $\hat{\sigma}_z$. 
This is a pure dephasing model 
because the interaction Hamiltonian 
$\hat{H}_{\rm int} = (\hbar/2)\hat{\xi} \hat{\sigma}_z$
between the qubit and bath includes only $\hat{\sigma}_z$.
In other words, there is no energy relaxation in this model 
($T_1$ is infinite in NMR terminology).
The spin (qubit) is subject to a static magnetic field $\Omega$ 
and a random magnetic field (noise) $\hat{\xi}$ in the $z$-direction; 
the random field $\hat{\xi}$ is generated by the bath.

There are several variations of the model 
depending on the nature of the bath.
Here we consider three of them. 
The first is a classical-noise model \cite{CywinskiDasSarma_etal}, 
where $\xi$ is a stationary Gaussian stochastic process 
(classical random variable) with zero mean.
(In this case, $\hat{H}_{\rm bath}$ does not appear.)
The second one is a spin-boson model \cite{ViolaLloyd, Uhrig, Uhrig2} 
in which $\hat{\xi} = \sum_j \lambda_j (\hat{b}_j^\dag + \hat{b}_j)$ 
and $\hat{H}_{\rm bath} = \sum_j \hbar \omega_j \hat{b}_j^\dag \hat{b}_j$, 
where $\hat{b}_j$ ($\hat{b}^\dag_j$) is the $j$th mode 
annihilation (creation) operator of the bosonic bath, 
and $\lambda_j$ is the coupling strength between the qubit and the $j$th mode boson.
The third one is a spin-spin bath model \cite{RaoKurizki} 
in which $\hat{\xi} = \sum_j \mu_j (\hat{s}_+^j + \hat{s}_-^j)$ 
and $\hat{H}_{\rm bath} = \sum_j (\hbar/2) \omega_j \hat{s}_z^j$, 
where $\hat{s}_+^j = \hat{s}_x^j + i \hat{s}_y^j$ and 
$\hat{s}_-^j = \hat{s}_x^j -i \hat{s}_y^j$. 
Here, $\hat{s}_x^j$, $\hat{s}_y^j$, and $\hat{s}_z^j$ are respectively 
the $x,y$, and $z$ components of the Pauli matrices of the $j$th spin in the bath, 
and $\mu_j$ is the coupling strength 
between the qubit (spin of interest) and the $j$th spin (in the bath).

The noise spectrum is defined as 
the symmetrized power spectral density function of the random field: 
\begin{align}
S (\omega) = \int_{-\infty}^\infty {\rm d}t ~ e^{i\omega t} 
\frac{1}{2} \Bigl\langle \Check{\xi}(t) \Check{\xi}(0) 
+ \Check{\xi}(0) \Check{\xi}(t) \Bigr\rangle_{\rm bath}.
\label{noiseSpectrum}
\end{align}
In the classical-noise model, 
$\langle \cdot \rangle_{\rm bath}$ is the average 
over the stochastic variable ($\Check{\xi}(t) = \xi(t)$).
In the spin-boson and spin-spin bath models, 
$\Check{\xi}(t) = e^{i \hat{H}_{\rm bath} t / \hbar} 
\hat{\xi} e^{-i \hat{H}_{\rm bath} t / \hbar}$, 
and $\langle \cdot \rangle_{\rm bath} = 
{\rm Tr}_{\rm bath} \bigl( \hat{\rho}_{\rm bath}^{\rm eq} \cdot \bigr)$ 
is the average in an equilibrium state of the bath 
($\hat{\rho}_{\rm bath}^{\rm eq} = e^{-\beta \hat{H}_{\rm bath}} / Z_{\rm bath}$).

{\it Pulse sequence and generalized coherence time}.---The density matrix 
(which describes the state) of the qubit is denoted by $\hat{\rho}_s$.
In the classical-noise model, $\hat{\rho}_s$ is given 
by the average of $\hat{\hat{\rho}}_s$ over the stochastic variable: 
$\hat{\rho}_s = \bigl\langle \hat{\hat{\rho}}_s \bigr\rangle_{\rm bath}$, 
where $\hat{\hat{\rho}}_s$ is the qubit state with one realization 
of the stochastic variable.
In the spin-boson and spin-spin bath models, $\hat{\rho}_s$ is given 
by the partial trace of the density matrix $\hat{\rho}$ of the total system: 
$\hat{\rho}_s = {\rm Tr}_{\rm bath} \hat{\rho}$.

The coherence is quantified using a non-diagonal component 
of the density matrix of the qubit: $\rho_{s,+-} = \langle + | \hat{\rho}_s | - \rangle$.
Here, $| \pm \rangle$ is the eigenstate of $\hat{\sigma}_z$
corresponding to the eigenvalue $\pm 1$.
Experimentally, the coherence is measured using 
the transverse ($x$ and $y$) components of the qubit (spin) 
and the relations 
$\langle \hat{\sigma}_x \rangle = 2{\rm Re} \rho_{s,+-}$ 
and $\langle \hat{\sigma}_y \rangle = -2{\rm Im} \rho_{s,+-}$.

\begin{figure}[tb]
\begin{center}
\includegraphics[width=.95\linewidth]{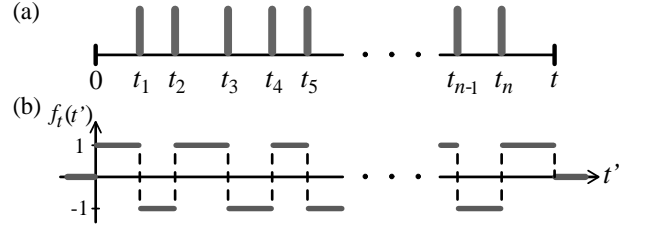}
\caption{(a) Illustration of a $\pi$-pulse sequence.
(b) Function $f_t(t')$ corresponding to illustration ($n$: even).}
\label{pulseSequence}
\end{center}
\end{figure}

Now we consider a situation in which ideal (instantaneous) $\pi$-pulses 
(about the $x$ or $y$ axis) are repeatedly applied to the qubit.
As shown in Fig.~\ref{pulseSequence}(a), 
the pulses are applied at times $t_1, t_2, ... , t_n$, 
where $n$ is the total number of pulses.
After the application of the sequential pulses, 
we measure the normalized coherence at time $t$ ($>t_n$): 
$W(t) = |\rho_{s,+-}(t)| / |\rho_{s,+-}(0)|$.

When we apply a sufficiently large number of pulses 
(keeping the minimal inter-pulse time fixed, so that $t$ is also sufficiently large), 
the coherence exhibits an exponential decay (as shown in the models later): 
\begin{align}
W(t) \sim \exp \left( -t / T_2^L \right).
\label{def_T2L}
\end{align}
This time dependence enables us to define uniquely the coherence time $T_2^L$ 
for a multiple-pulse sequence, which we call ``generalized'' coherence time
(for a reason mentioned later).
$T_2^L$ differs from conventional coherence time $T_2^{\rm SE}$, 
which is defined using spin echo (SE) experiments 
($T_2^L > T_2^{\rm SE}$ in most cases). 
Generally, it can be shown that $T_2^L$ depends on the pulse sequence.

In the three models, 
the time evolution of the normalized coherence 
in the presence of a pulse sequence can be expressed as 
(weak coupling condition $\mu_j \ll \omega_j$ is necessary for the spin-spin bath model)
\cite{Uhrig, Uhrig2, CywinskiDasSarma_etal, RaoKurizki, note_W}
\begin{align}
W(t) &= \exp \left( -\int_0^\infty \frac{{\rm d}\omega}{2\pi} 
S(\omega) \big| \tilde{f}_t(\omega) \big| ^2 \right),
\label{signal_S}
\\
\tilde{f}_t(\omega) &= \int_{-\infty}^\infty {\rm d}t' e^{i\omega t'} f_t(t'),
\label{FT_f_t}
\\
f_t(t') &= \displaystyle \sum_{k=0}^n  (-1)^k \Theta(t_{k+1} - t') \Theta(t' - t_k),
\label{f_t}
\end{align}
where $t_0=0$, $t_{n+1}=t$, and 
$\{ t_k \}_{k=1}^n$ is the set of times when the pulses are applied, 
and $\Theta(T)$ is the Heaviside step function.
As shown in Fig.~\ref{pulseSequence}(b), 
$f_t(t')$ depends on the pulse sequence and 
takes a non-zero value  ($+1$ or $-1$) only for $0<t'<t$.

{\it Relationship between $T_2^{\rm SE}$ and $S$}.---Before presenting the main results, 
we show the relationship between $T_2^{\rm SE}$ and $S$.
For the SE pulse sequence ($\pi/2(x)$-$\tau$-$\pi(x)$-$\tau$-signal), 
%
%
we can easily find that $| \tilde{f}_t(\omega) |^2 = (16/\omega^2) \sin^4 (\omega\tau/2)$.
Substituting this into Eq.~(\ref{signal_S}) and using the asymptotic behavior 
$(1/\omega^2) \sin^4 (\omega\tau/2) \sim (\pi/4) \delta(\omega) \tau$
as $\tau \to \infty$, 
we get the asymptotic $\tau$-dependence of the SE coherence $W^{\rm SE}(2\tau)$, 
\begin{align}
W^{\rm SE}(2\tau) \sim \exp \left( - \frac{1}{2} S(0) 2\tau \right) 
~~\text{as } \tau\to\infty, 
\label{WSE}
\end{align}
which yields the relationship between $T_2^{\rm SE}$ and $S$: 
\begin{align}
\frac{1}{T_2^{\rm SE}} = \frac{1}{2} S(0).
\label{T2SE_final}
\end{align}
Although this formula is common in the field of NMR \cite{Slichter, Recchia},
it has a significant implication; 
we should use the asymptotic exponential decay 
of the SE coherence to define $T_2^{\rm SE}$, 
as seen in Eq.~(\ref{WSE}) \cite{Fine, note_SE}.
This is contrastive to the usual evaluation of coherence time in SE experiments, 
in which the behaviors of the SE coherence at smaller $\tau$ are generally used. 
Their use would not provide a unique definition of coherence time 
because the functional form of the SE coherence at smaller $\tau$ 
depends on the systems. 
[In some systems exponential decays are observed 
with certain time constants (different from $T_2^{\rm SE}$ in some cases), 
while in other systems Gaussian decays are observed 
with certain time constants (different from $T_2^{\rm SE}$).]

{\it Relationship between $T_2^L$ and $S$}.---In the long time limit 
(large $n$ limit) keeping the minimal inter-pulse time fixed, 
we get the power spectral density function of $f_t(t')$:
$I(\omega) \equiv \lim_{t \to \infty}(1/t) \big| \tilde{f}_t(\omega) \big| ^2$. 
Hence $\big| \tilde{f}_t(\omega) \big| ^2 \sim I(\omega)t$ 
as $t \to \infty$.
From this and Eq.~(\ref{signal_S}), 
we find that the asymptotic behavior of the coherence is an exponential decay: 
\begin{align}
W(t) \sim \exp \left( -\int_0^\infty \frac{{\rm d}\omega}{2\pi} 
S(\omega) I(\omega) t \right) ~~ \text{as } t \to \infty.
\label{signal_I}
\end{align}
The necessary number of pulses to extract this asymptotic exponential decay 
is independent of the noise spectrum because $f_t(t')$ is determined 
only by the pulse sequence.
Comparing this equation with the definition of $T_2^L$ [Eq.~(\ref{def_T2L})],
we obtain 
\begin{align}
\frac{1}{T_2^L} 
= \int_0^\infty \frac{{\rm d}\omega}{2\pi} S(\omega) I(\omega).
\label{T2L_S}
\end{align}

If the sequence consists of equidistant pulses 
with $t_{k+1} - t_k = 2\tau$ ($k = 1,2,...,n-1$), 
$f_t(t')$ is a periodic function of $t'$ with a period of $4\tau$ (for $0<t'<t$).
In this case, the power spectral density function $I(\omega)$ 
(with fixed $\tau$) becomes 
\begin{align}
I(\omega) = 2\pi \sum_{m=-\infty}^\infty \big| C_m \big|^2 \delta(\omega - \omega_m),
\label{I_peri}
\end{align}
where $\omega_m = m\pi/2\tau$, and 
$C_m$ is the Fourier coefficient of $f_t(t')$: 
$C_m = (1/4\tau)\int_0^{4\tau} {\rm d}t' ~e^{i \omega_m t'} f_t(t')$.
Substituting Eq.~(\ref{I_peri}) into Eq.~(\ref{T2L_S}), we obtain
\begin{align}
\frac{1}{T_2^L} = \sum_{m=0}^\infty \big| C_m \big|^2 S(\omega_m).
\end{align}

As an example, we investigate the use of the APCP pulse sequence: 
$\pi/2(x)$-$\{\tau$-$\pi(x)$-$2\tau$-$\pi(\bar{x})$-$\tau\}^{n/2}$-signal ($n$: even).
For this pulse sequence, 
\begin{align}
f_t(t') = 
\begin{cases}
1 & \text{for } 4k\tau <t' < (4k+1)\tau 
\\
& \text{and } (4k+3)\tau <t' < 4(k+1)\tau ,
\\
-1& \text{for } (4k+1)\tau <t' < (4k+3)\tau , 
\\
0 & \text{otherwise},
\end{cases}
\notag
\end{align}
where $k=0,1,2,...,n-1$.
This yields $\big| C_m \big|^2 = (4/\pi^2 m^2)\delta_{m,2l+1}$ ($l=0,1,2,...$). 
Hence we finally obtain 
\begin{align}
\frac{1}{T_2^L} = \frac{4}{\pi^2}\sum_{l=0}^\infty \frac{1}{(2l+1)^2} S(\omega_{2l+1}).
\label{T2L_final}
\end{align}
Note that we obtain the same result 
for the Carr-Purcell and CPMG sequences   
because we assume ideal (instantaneous) pulses.
[To avoid pulse-error accumulation in actual experiments (with non-ideal pulses), 
one should use CPMG or APCP.]
Since the factor $1/(2l+1)^2$ is smaller 
for larger $l$,
we can approximate the above equation into
\begin{align}
\frac{1}{T_2^L} \simeq \frac{4}{\pi^2} S \bigl( \pi/2\tau \bigr),
\label{T2L_app}
\end{align}
if $S(\omega)$ rapidly decreases as $\omega$ increases.
Equations (\ref{T2L_final}) and (\ref{T2L_app}) are the main results of this Letter.

These relationships are qualitatively explained as follows.
In many systems, coherence time is dominated 
by the lowest-frequency component of the noise spectrum
(fluctuation-dissipation relation).
The pulse sequence with time interval $\sim\tau$ cancels out the noise 
at frequencies lower than $1/\tau$ (dynamical decoupling).
Therefore, the noise spectrum around the frequency $1/\tau$ 
dominantly contributes to the coherence time 
in the presence of the pulse sequence.

Note that the infinite $\tau$ limit of Eq.~(\ref{T2L_final}) is consistent 
with Eq.~(\ref{T2SE_final}): 
\begin{align}
\lim_{\tau\to\infty} \frac{1}{T_2^L} 
&= \frac{4}{\pi^2} S(0) \sum_{l=0}^\infty \frac{1}{(2l+1)^2}
\notag\\
&= \frac{1}{2} S(0)
= \frac{1}{T_2^{\rm SE}},
\label{T2L_T2SE}
\end{align}
where we use $\lim_{\tau\to\infty} \omega_{2l+1} = 0, \forall l$ in the first line, 
and $\sum_{l=0}^\infty 1/(2l+1)^2 = \pi^2  / 8$ in the second line.
This equation enables us to interpret $T_2^L$ 
as a generalization of $T_2^{\rm SE}$ into non-zero frequencies, 
which is the reason we call $T_2^L$ ``generalized coherence time.''

Similarly, the infinitesimal $\tau$ limit of Eq.~(\ref{T2L_final}) is
given by
\begin{align}
\lim_{\tau \to +0} \frac{1}{T_2^L} 
%
%
&= \frac{1}{2} \lim_{\omega\to\infty} S(\omega).
\end{align}

{\it Measurement of noise spectrum}.---Using the approximate 
relation (\ref{T2L_app}) we evaluate the noise spectrum as follows.
For a fixed value of $2\tau$ (inter-pulse time), 
we evaluate $T_2^L$ by applying a sufficiently large number of pulses 
and measuring the asymptotic behavior of the coherence.
We repeat this procedure for other fixed values of $2\tau$.
Then, by plotting $1/T_2^L$ against $\pi/2\tau$, 
we obtain $S$ as a function of $\omega$.

Better evaluation of the spectrum is possible 
by using the precise relation (\ref{T2L_final}).
For example, a functional form $S^{\rm fit}(\omega)$ of the noise spectrum 
(which includes some fitting parameters) based on the above approximate evaluation
is phenomenologically introduced.
Then we create a function $F(\pi/2\tau)$ 
similar to that on the right-hand side of Eq.~(\ref{T2L_final}) 
by summing the phenomenological function $S^{\rm fit}$
up to an appropriate cutoff $L$ ($L \gtrsim 10$ would be sufficient):
\begin{align}
F(\pi/2\tau) = \frac{4}{\pi^2}\sum_{l=0}^L \frac{1}{(2l+1)^2} 
S^{\rm fit} \bigl( (2l+1)\pi/2\tau \bigr) .
\end{align}
We finally fit $F(\pi/2\tau)$ to the experimental results 
($1/T_2^L$ vs $\pi/2\tau$) 
to obtain the values of the parameters for $S^{\rm fit}$.

Finally, we estimate the frequency range for this method of noise measurement.
To measure experimentally the asymptotic behavior of the coherence, 
we should apply a large number of pulses 
before the amplitude becomes too small to detect.
Hence, the inter-pulse time $2\tau$ must be smaller 
than the $1/e$ decay time $T_2$ of the coherence 
(in the presence of the SE pulse sequence).
This implies that the lower bound on the frequency $\pi/2\tau$ should be $\pi/T_2$.
The upper bound is determined by 
the shortest inter-pulse time that is experimentally available.
The value of this time can be of the same order as 
that of the pulse duration time $\tau_p$.
Hence, the upper bound on the frequency $\pi/2\tau$ should be $\pi/\tau_p$.

{\it Concluding remarks}.---In summary, we have described 
a method for obtaining the dephasing noise spectrum.
The method is simple in the sense that
we have only to apply sequences of equidistant $\pi$-pulses to the qubit (spin). 
The generalized coherence time, evaluated from the asymptotic exponential decay 
of the coherence in the presence of a sufficiently large number of pulses, 
is used for evaluating the spectrum. 
This method is applicable to a system 
interacting with several independent noise sources. 
In this case, the total spectrum is the sum of the individual noise spectra.

This method is expected to be valid in a wide range of systems 
because we have derived it for three models.
The single-qubit noise spectrum plays an important role 
even in multi-qubit systems 
because it significantly contributes to the dynamics of the systems
(this is clearly seen when analyzing with the quantum master equation).
Extension of the present study remains a theoretical task.
We should analyze a model 
in which the system-environment coupling is given in a general form 
and/or in which there are both energy and phase relaxations.
The projection operator formalism \cite{GordonKurizki_etal} would be helpful
in analyzing a general model.

This method should provide new insights 
into NMR experiments in condensed matter physics.
So far the longitudinal relaxation time $T_1$ has been 
successfully used for investigating properties of electrons 
and nuclear spins in condensed matter.
The method presented here should enable the use of the  
generalized coherence time $T_2^L$ for investigating (other) properties of them.
This is because the noise spectrum (evaluated from $T_2^L$) 
includes information on the environment, the physical origin of which is 
electrons, other nuclear spins, and so on.
In order to make this application useful, 
it is important to capture the characteristic structure of the spectrum
by analyzing microscopic models that include interactions with nuclear spins 
(e.g., Fermi contact hyperfine and dipolar couplings).
An experimental demonstration of the present method in NMR 
will be reported elsewhere \cite{Sasaki_etal}.

\begin{acknowledgments}
The authors thank K. Akiba and S. Watanabe for helpful discussions.
This work was supported by a Grant-in-Aid for the GCOE Program 
``Weaving Science Web beyond Particle-Matter Hierarchy.''
\end{acknowledgments}

\clearpage

\setcounter{section}{1}
\setcounter{equation}{0}
\renewcommand{\thesection}{\Alph{section}}
\numberwithin{equation}{section}

\section{Supplemental Material}

In this supplement, we give detailed derivations of Eq.~(\ref{signal_S}) 
for the three models for completeness.
The essentials of the derivations are the same as those 
in Refs.~\cite{Uhrig, Uhrig2, CywinskiDasSarma_etal, RaoKurizki}.

\subsection{Classical-noise model}

The Hamiltonian of the classical noise model is 
\begin{align}
\hat{H}(t) = \frac{\hbar}{2} \bigl( \Omega + \xi(t) \bigr) \hat{\sigma}_z, 
\label{H_classical}
\end{align}
where $\xi(t)$ is a stationary stochastic process with zero mean. 
In this model the noise spectrum defined by Eq.~(\ref{noiseSpectrum}) 
can be rewritten as 
\begin{align}
S (\omega) = \int_{-\infty}^\infty {\rm d}t ~ 
e^{i\omega t} \langle \xi(t) \xi(0) \rangle_{\rm bath}.
\end{align}

The relation between the states at times $t_k+0$ (immediately after the $k$th pulse) 
and $t_{k+1}-0$ (immediately before the $(k+1)$th pulse) is  
\begin{align}
\hat{\hat{\rho}}_s (t_{k+1}-0) 
= \hat{U}(t_{k+1},t_k) \hat{\hat{\rho}}_s (t_k+0) \hat{U}(t_{k+1},t_k)^\dag,
\label{classical_k+1_k}
\end{align}
where the unitary time evolution operator $\hat{U}$ is given by 
\begin{align}
\hat{U}(t_{k+1},t_k) &= 
\exp \left[ -\frac{i}{\hbar} \int_{t_k}^{t_{k+1}} {\rm d}t' \hat{H}(t') \right]
\notag\\
&= \exp\left( -\frac{i}{2} \hat{\sigma}_z \theta_k \right).
\notag
\end{align}
Here, $\theta_k = \Omega (t_{k+1} - t_k) + \int_{t_k}^{t_{k+1}} {\rm d}t' \xi(t')$.
At time $t_{k+1}$, an ideal $\pi$-pulse is applied about the $x$ ($y$) axis, so 
\begin{align}
\hat{\hat{\rho}}_s (t_{k+1}+0) 
= \hat{D}_{x(y)} (\pm \pi) \hat{\hat{\rho}}_s (t_{k+1}-0) \hat{D}_{x(y)} (\pm \pi)^\dag , 
\label{classical_k+1_k+1}
\end{align}
where the $\pm\pi$-rotation operator about the $x$ ($y$) axis is 
$\hat{D}_{x(y)}(\pm \pi) = \mp i \hat{\sigma}_{x(y)}$.

Substituting Eq.~(\ref{classical_k+1_k}) into Eq.~(\ref{classical_k+1_k+1}), 
we obtain 
\begin{align}
\langle + | \hat{\hat{\rho}}_s  (t_{k+1}+0) | - \rangle
&= \pm \langle - | \hat{\hat{\rho}}_s  (t_k+0) | + \rangle e^{i\theta_k},
\\
\langle - | \hat{\hat{\rho}}_s  (t_{k+1}+0) | + \rangle
&= \pm \langle + | \hat{\hat{\rho}}_s  (t_k+0) | - \rangle e^{-i\theta_k}.
\end{align}
Here, the upper (lower) signs are for the rotation about the $x$ ($y$) axis.
By solving these recurrence relations, we obtain 
\begin{align}
\langle & + | \hat{\hat{\rho}}_s  (t_n+0) | - \rangle
\notag\\
&= (-1)^{n_y} \langle \pm | \hat{\hat{\rho}}_s  (0) | \mp \rangle
\exp\left[ -i \sum_{k=0}^{n-1} (-1)^{k+n} \theta_k \right],
\notag
\end{align}
where the upper (lower) signs are for even (odd) $n$, 
and $n_y$ is the total number of rotations about the $y$ axis.
Then, from the state at time $t$ (at which the signal is measured) 
$\hat{\hat{\rho}}_s (t) = \hat{U}(t,t_n) \hat{\hat{\rho}}_s (t_n+0) \hat{U}(t,t_n)^\dag$, 
we obtain 
\begin{align}
\langle & + | \hat{\hat{\rho}}_s (t) | - \rangle
\notag\\
&= (-1)^{n_y} \langle \pm | \hat{\hat{\rho}}_s (0) | \mp \rangle
\exp\left[ -i \sum_{k=0}^n (-1)^{k+n} \theta_k \right],
\notag
\end{align}

\begin{widetext}
Averaging over the stochastic variables, we obtain 
\begin{align}
\rho_{s,+-} (t) 
&= \bigl\langle + \big| 
\bigl\langle \hat{\hat{\rho}}_s (t)\bigr\rangle_{\rm bath} \big| - \bigr\rangle 
\notag\\
&= (-1)^{n_y} \rho_{s,\pm \mp} (0) 
\exp\left[ -i \Omega \sum_{k=0}^n (-1)^{k+n} (t_{k+1}-t_k) \right]
\left\langle \exp\left[ -i \sum_{k=0}^n (-1)^{k+n} 
\int_{t_k}^{t_{k+1}} {\rm d}t' \xi(t') \right] \right\rangle_{\rm bath}.
\notag
\end{align}
The normalized coherence $W(t)$
is then given by (note that $|\rho_{s,+-}| = |\rho_{s,-+}|$)
\begin{align}
W(t) &= \Bigg|\left\langle \exp\left[ -i \sum_{k=0}^n (-1)^{k+n} 
\int_{t_k}^{t_{k+1}} {\rm d}t' \xi(t') \right] \right\rangle_{\rm bath} \Bigg|
\notag\\
&= \Bigg|\left\langle \exp\left[ i (-1)^{n+1} 
\int_0^t {\rm d}t' \xi(t') f_t(t') \right] \right\rangle_{\rm bath} \Bigg|,
\end{align}
where $f_t(t')$ is defined by Eq.~(\ref{f_t}). 

If $\xi(t)$ is a stationary Gaussian process
we obtain the desired result: 
\begin{align}
W(t) &= \Bigg| \exp\left[ - \frac{1}{2}
\int_0^t {\rm d}t'_1 \int_0^t {\rm d}t'_2 \langle \xi(t'_1) 
\xi(t'_2) \rangle_{\rm bath} 
f_t(t'_1) f_t(t'_2) \right] \Bigg|
\notag\\
&= \exp\left[ - \int_0^\infty \frac{{\rm d}\omega}{2\pi}
S(\omega) \big| \tilde{f}_t(\omega) \big|^2 \right] .
\end{align}
\end{widetext}


\subsection{Analysis of pure dephasing model with quantum bath}

Here we analyze a pure dephasing model 
described by the Hamiltonian (\ref{hamiltonian}).
We assume that the Hamiltonian can be decoupled into the qubit and bath parts 
by a unitary transformation: 
\begin{align}
\hat{H}^e &= e^{\hat{\sigma}_z \hat{K}} \hat{H} e^{-\hat{\sigma}_z \hat{K}} 
\notag\\
&= \frac{\hbar\Omega}{2}\hat{\sigma}_z + \hat{H}_{\rm bath}^e + (\text{const}.),
\label{decouple}
\end{align}
where $\hat{K}$ and $\hat{H}_{\rm bath}^e$ are respectively anti-Hermitian 
and Hermitian operators of the bath. 
Using the same unitary transformation, 
we obtain effective qubit operators and effective time evolution operators:
\begin{align}
\hat{\sigma}_{x(y)}^e 
&= e^{\hat{\sigma}_z \hat{K}} \hat{\sigma}_{x(y)} e^{-\hat{\sigma}_z \hat{K}} 
\notag\\
&= e^{2\hat{\sigma}_z \hat{K}} \hat{\sigma}_{x(y)} 
= \hat{\sigma}_{x(y)} e^{-2\hat{\sigma}_z \hat{K}},
\\
\hat{U}^e(t_{k+1},t_k) 
&= e^{\hat{\sigma}_z \hat{K}} \exp\bigl[-i\hat{H}(t_{k+1}-t_k)/\hbar\bigr] 
e^{-\hat{\sigma}_z \hat{K}}
\notag\\
&= \exp\bigl[-i\hat{H}^e(t_{k+1}-t_k)/\hbar\bigr] .
\end{align}

\begin{widetext}
The relation between the states 
at times $t_k+0$ (immediately after the $k$th pulse) 
and $t_{k+1}-0$ (immediately before the $(k+1)$th pulse) is  
\begin{align}
\hat{\rho}(t_{k+1}-0) 
= \exp\bigl[-i\hat{H}(t_{k+1}-t_k)/\hbar\bigr] \hat{\rho}(t_k+0) 
\exp\bigl[i\hat{H}(t_{k+1}-t_k)/\hbar\bigr].
\label{quantum_k+1_k}
\end{align}
At time $t_{k+1}$, an ideal $\pi$-pulse is applied about the $x$ ($y$) axis, so 
\begin{align}
\hat{\rho}(t_{k+1}+0) 
= \hat{D}_{x(y)} (\pm \pi) \hat{\rho}(t_{k+1}-0) \hat{D}_{x(y)} (\pm \pi)^\dag , 
\label{quantum_k+1_k+1}
\end{align}
where the $\pm\pi$-rotation operator about the $x$ ($y$) axis is 
$\hat{D}_{x(y)}(\pm \pi) = \mp i \hat{\sigma}_{x(y)}$.

Substituting Eq.~(\ref{quantum_k+1_k}) into Eq.~(\ref{quantum_k+1_k+1}), 
we obtain 
\begin{align}
\hat{\rho} (t_{k+1}+0) 
&= \hat{\sigma}_{x(y)} \exp\bigl[-i\hat{H}(t_{k+1}-t_k)/\hbar\bigr] \hat{\rho} (t_k+0) 
\exp\bigl[i\hat{H}(t_{k+1}-t_k)/\hbar\bigr] \hat{\sigma}_{x(y)},
\notag\\
\hat{\rho}^e (t_{k+1}+0) 
&\equiv e^{\hat{\sigma}_z \hat{K}} \hat{\rho} (t_{k+1}+0) e^{-\hat{\sigma}_z \hat{K}}
\notag\\
&= e^{2\hat{\sigma}_z \hat{K}} \hat{\sigma}_{x(y)} \hat{U}^e(t_{k+1},t_k) 
\hat{\rho}^e (t_k+0) 
\hat{U}^e(t_{k+1},t_k)^\dag \hat{\sigma}_{x(y)} e^{-2\hat{\sigma}_z \hat{K}},
%
\notag\\
\langle + | \hat{\rho}(t_{k+1}+0) | - \rangle
&= \langle + | e^{-\hat{\sigma}_z \hat{K}} \hat{\rho}^e (t_{k+1}+0) 
e^{\hat{\sigma}_z \hat{K}} | - \rangle
\notag\\
&= \pm e^{i\Omega (t_{k+1}-t_k)} e^{\hat{K}} 
e^{-i\hat{H}_{\rm bath}^e(t_{k+1}-t_k)/\hbar} e^{-\hat{K}} 
\langle - | \hat{\rho} (t_k+0) | + \rangle e^{-\hat{K}}
e^{i\hat{H}_{\rm bath}^e(t_{k+1}-t_k)/\hbar} e^{\hat{K}} 
\notag\\
&= \pm e^{i\Omega (t_{k+1}-t_k)} \hat{T}^{\rm bath+}_{k+1,k} 
\langle - | \hat{\rho} (t_k+0) | + \rangle \hat{T}^{\rm bath+}_{k+1,k}{}^\dag , 
\label{recurrence1}
\\
\langle - | \hat{\rho}(t_{k+1}+0) | + \rangle
&= \pm e^{-i\Omega (t_{k+1}-t_k)} e^{-\hat{K}} 
e^{-i\hat{H}_{\rm bath}^e(t_{k+1}-t_k)/\hbar} e^{\hat{K}} 
\langle - | \hat{\rho} (t_k+0) | + \rangle e^{\hat{K}}
e^{i\hat{H}_{\rm bath}^e(t_{k+1}-t_k)/\hbar} e^{-\hat{K}} 
\notag\\
&= \pm e^{-i\Omega (t_{k+1}-t_k)} \hat{T}^{\rm bath-}_{k+1,k} 
\langle + | \hat{\rho} (t_k+0) | - \rangle \hat{T}^{\rm bath-}_{k+1,k}{}^\dag .
\label{recurrence2}
\end{align}
Here, $\hat{T}^{\rm bath+}_{k+1,k} 
= e^{\hat{K}} e^{-i\hat{H}_{\rm bath}^e(t_{k+1}-t_k)/\hbar} e^{-\hat{K}}$, and 
$\hat{T}^{\rm bath-}_{k+1,k} 
= e^{-\hat{K}} e^{-i\hat{H}_{\rm bath}^e(t_{k+1}-t_k)/\hbar} e^{\hat{K}}$; 
the upper (lower) signs are for rotation about the $x$ ($y$) axis.

By solving these recurrence relations (\ref{recurrence1}) and (\ref{recurrence2}), 
we obtain 
\begin{align}
\langle + | \hat{\rho}(t_n+0) | - \rangle
=& (-1)^{n_y} e^{-i \sum_{k=0}^{n-1} (-1)^{k+n} \Omega (t_{k+1}-t_k)} 
\hat{T}^{\rm bath+}_{n,n-1} \hat{T}^{\rm bath-}_{n-1,n-2} \cdots \hat{T}^{\rm bath \mp}_{1,0} 
\notag\\
&\times \langle \pm | \hat{\rho} (0) | \mp \rangle \hat{T}^{\rm bath \mp}_{1,0}{}^\dag 
\cdots \hat{T}^{\rm bath-}_{n-1,n-2}{}^\dag \hat{T}^{\rm bath+}_{n,n-1}{}^\dag,
\label{quantum_t_n}
\end{align}
where the upper (lower) signs are for even (odd) $n$, 
and $n_y$ is the total number of rotations about the $y$ axis.
From the state at time $t$ (when the signal is measured)  
$\hat{\rho}(t) = \exp\bigl[-i\hat{H}(t-t_n)/\hbar\bigr] 
\hat{\rho}(t_n+0) \exp\bigl[i\hat{H}(t-t_n)/\hbar\bigr]$, 
we have 
\begin{align}
\langle + | \hat{\rho}(t) | - \rangle 
&= \langle + | \exp\bigl[-i\hat{H}(t-t_n)/\hbar\bigr] | + \rangle 
\langle + | \hat{\rho}(t_n+0) | - \rangle 
\langle - | \exp\bigl[-i\hat{H}(t-t_n)/\hbar\bigr] | - \rangle 
\notag\\
&= e^{-i\Omega (t-t_n)} e^{-\hat{K}} e^{-i\hat{H}_{\rm bath}^e(t-t_n)/\hbar} e^{\hat{K}}
\langle + | \hat{\rho}(t_n+0) | - \rangle 
e^{\hat{K}} e^{i\hat{H}_{\rm bath}^e(t-t_n)/\hbar} e^{-\hat{K}},
\label{quantum_t}
\end{align}
where we have used 
$\langle - | \exp\bigl[-i\hat{H}(t-t_n)/\hbar\bigr] | + \rangle = 
\langle + | \exp\bigl[-i\hat{H}(t-t_n)/\hbar\bigr] | - \rangle = 0$.
Combining Eqs.~(\ref{quantum_t_n}) and (\ref{quantum_t}) and 
taking the partial trace of the bath, we obtain 
\begin{align}
\rho_{+-} (t) &= {\rm Tr}_{\rm bath} \langle + | \hat{\rho}(t) | - \rangle
\notag\\
&= (-1)^{n_y} e^{-i \sum_{k=0}^n (-1)^{k+n} \Omega (t_{k+1}-t_k)} 
{\rm Tr}_{\rm bath} \Bigl[ \langle \pm | \hat{\rho}(0) 
| \mp \rangle e^{(-1)^n 2 \hat{L}} \Bigr],
\end{align}
where $t_0=0$, $t_{n+1}=t$, and 
\begin{align}
\hat{L} = \hat{K} + \sum_{k=0}^n (-1)^k 2\Check{K}(t_k) +(-1)^{n+1}\Check{K}(t).
\label{L}
\end{align}
Here, $\Check{K}(T) = e^{i\hat{H}_{\rm bath}^e T/\hbar} \hat{K} 
e^{-i\hat{H}_{\rm bath}^e T/\hbar}$.
If the initial state is a decoupled one, 
$\hat{\rho}(0) = \hat{\rho}_s(0) \hat{\rho}_{\rm bath}^{\rm eq}$, 
\begin{align}
\rho_{+-} (t) 
&= (-1)^{n_y} e^{-i \sum_{k=0}^n (-1)^{k+n} \Omega (t_{k+1}-t_k)} 
{\rm Tr}_{\rm bath} \Bigl[ \hat{\rho}_{\rm bath}^{\rm eq} 
e^{(-1)^n 2 \hat{L}} \Bigr] \rho_{\pm \mp} (0) ,  
\\
W(t) &= \Big| {\rm Tr}_{\rm bath} \Bigl[ \hat{\rho}_{\rm bath}^{\rm eq} 
e^{(-1)^n 2 \hat{L}} \Bigr] \Big|.
\label{coherence_quantum}
\end{align}
\end{widetext}

\subsection{Spin-boson model}
The Hamiltonian of the spin-boson model is 
\begin{align}
\hat{H} = 
\frac{\hbar}{2} \Bigl( \Omega + \hat{\xi} \Bigr) \hat{\sigma}_z + \hat{H}_{\rm bath} , 
\end{align}
where $\hat{\xi} = \sum_j \lambda_j (\hat{b}_j^\dag + \hat{b}_j)$ 
and $\hat{H}_{\rm bath} = \sum_j \hbar \omega_j \hat{b}_j^\dag \hat{b}_j$.
The {\it spectral density of the bath} $J(\omega)$ is defined by 
\begin{align}
J(\omega) = \sum_j \lambda_j^2 \delta(\omega - \omega_j),
\end{align}
for $\omega \ge 0$.
The relation between $J(\omega)$ and $S(\omega)$ is 
\begin{align}
S(\omega) 
=  \pi J( | \omega | ) \bigl( 2n_b( | \omega | ) + 1 \bigr) .
\label{S_J_spinboson}
\end{align}
Here, $n_b (\omega) = 1/(e^{\beta\hbar\omega} -1)$ is the boson occupation factor.

The spin-boson model satisfies condition ({\ref{decouple}) 
if we set $\hat{K} = \sum_j \lambda_j (\hat{b}_j^\dag - \hat{b}_j)/2\omega_j$: 
\begin{align}
\hat{H}^e &= e^{\hat{\sigma}_z \hat{K}} \hat{H} e^{-\hat{\sigma}_z \hat{K}} 
\notag\\
&= \frac{\hbar\Omega}{2}\hat{\sigma}_z + \hat{H}_{\rm bath}^e + \Delta E,
\end{align}
where $\hat{H}_{\rm bath}^e = \hat{H}_{\rm bath} 
= \sum_j \hbar \omega_j \hat{b}_j^\dag \hat{b}_j$, 
and $\Delta E = -\int_0^\infty {\rm d}\omega J(\omega)/4\omega$.
In this case, Eq.~(\ref{L}) becomes 
\begin{align}
\hat{L} = -i \sum_j \frac{\lambda_j}{2} 
\Bigl[ \tilde{f}_t(\omega_j) \hat{b}_j^\dag - \tilde{f}^*_t(\omega_j) \hat{b}_j \Bigr], 
\end{align}
where $\tilde{f}_t(\omega)$ is given by Eq.~(\ref{FT_f_t}).
Then, by applying the Wick theorem, we transform Eq.~(\ref{coherence_quantum}) into 
\begin{align}
W(t) &= \exp \Bigl( 2 \bigl\langle \hat{L}^2 \bigr\rangle_{\rm bath} \Bigr)
\notag\\
&= \exp \left( -\frac{1}{2} \sum_j \lambda_j^2 \big| \tilde{f}_t(\omega_j) \big| ^2 
\bigl\langle \hat{b}_j^\dag \hat{b}_j + \hat{b}_j \hat{b}_j^\dag \bigr\rangle_{\rm bath} \right)
\notag\\
&= \exp \left( -\frac{1}{2}\int_0^\infty {\rm d}\omega 
J(\omega) \big| \tilde{f}_t(\omega) \big| ^2 
\bigl( 2n_b(\omega) +1 \bigr) \right).
\notag
\end{align}
Using Eq.~(\ref{S_J_spinboson}), we finally obtain the desired result (\ref{signal_S}).

\subsection{Spin-spin bath model}
The Hamiltonian of the spin-spin bath model is 
\begin{align}
\hat{H} = 
\frac{\hbar}{2} \Bigl( \Omega + \hat{\xi} \Bigr) \hat{\sigma}_z + \hat{H}_{\rm bath} , 
\end{align}
where $\hat{\xi} = \sum_j \mu_j (\hat{s}_+^j + \hat{s}_-^j)$ 
and $\hat{H}_{\rm bath} = \sum_j (\hbar/2) \omega_j \hat{s}_z^j$.
The {\it spectral density of the bath} $J(\omega)$ is defined by 
\begin{align}
J(\omega) = \sum_j \mu_j^2 \delta(\omega - \omega_j) , 
\end{align}
for $\omega \ge 0$.
The relation between $J(\omega)$ and $S (\omega)$ is 
\begin{align}
S (\omega) = 4\pi J( | \omega |) . 
\label{S_J_spinspin}
\end{align}

The spin-spin bath model satisfies condition ({\ref{decouple}) 
if we set $\hat{K} = (i/2) \sum_j \alpha_j  \hat{s}^j_y$: 
\begin{align}
\hat{H}^e &= e^{\hat{\sigma}_z \hat{K}} \hat{H} e^{-\hat{\sigma}_z \hat{K}} 
\notag\\
&= \frac{\hbar\Omega}{2}\hat{\sigma}_z + \hat{H}_{\rm bath}^e,
\end{align}
where $\hat{H}_{\rm bath}^e = 
(\hbar/2) \sum_j ( \omega_j \cos\alpha_j + 2\mu_j \sin\alpha_j ) \hat{s}^j_z$, 
and $\alpha_j = {\rm arctan} (2\mu_j/\omega_j)$. 
In this case, Eq.~(\ref{L}) becomes 
\begin{align}
\hat{L} = \frac{i}{2} \sum_j \alpha_j \gamma_j 
\left[ \hat{s}^j_x {\rm Re} \tilde{f}_t(\gamma_j) 
- \hat{s}^j_y {\rm Im} \tilde{f}_t(\gamma_j) \right],
\end{align}
where $\gamma_j = \sqrt{\omega_j^2 + 4\mu_j^2}$, and 
$\tilde{f}_t(\omega)$ is given by Eq.~(\ref{FT_f_t}).
Then 
\begin{align}
e^{(-1)^n 2\hat{L}} = \prod_j \bigl[ \cos\theta_j 
+ (-1)^n \sin\theta_j (X_j \hat{s}^j_x + Y_j \hat{s}^j_y ) \bigr],
\notag
\end{align}
where $\theta_j = \alpha_j \gamma_j \bigl|\tilde{f}_t(\gamma_j) \big|$, 
$X_j = {\rm Re} \tilde{f}_t(\gamma_j) / \bigl|\tilde{f}_t(\gamma_j) \big|$, 
and $Y_j = - {\rm Im} \tilde{f}_t(\gamma_j) / \bigl|\tilde{f}_t(\gamma_j) \big|$.
Therefore Eq.~(\ref{coherence_quantum}) becomes 
\begin{align}
W(t) &= \prod_j \big| \cos\theta_j \big|
\notag\\
&= \exp\left( \sum_j \ln\big| \cos\theta_j \big| \right).
\label{signalSSB}
\end{align}

If $\mu_j \ll \omega_j$ (the coupling between the qubit and the bath is weak), 
$\alpha_j = 2\mu_j / \omega_j + O\bigl( (\mu_j / \omega_j)^3 \bigr)$, 
$\gamma_j = \omega_j + O\bigl( (\mu_j / \omega_j)^2 \bigr)$, and 
$\tilde{f}_t(\gamma_j) = \tilde{f}_t(\omega_j) + O\bigl( (\mu_j / \omega_j)^2\bigr)$.
Therefore, $\theta_j = 2\mu_j \bigl|\tilde{f}_t(\omega_j) \big| 
+ O\bigl( (\mu_j / \omega_j)^3\bigr)$, and 
$\ln\big| \cos\theta_j \big| = -2 \mu_j^2 \bigl|\tilde{f}_t(\omega_j) \big|^2
+ O\bigl( (\mu_j / \omega_j)^3\bigr)$.
Substituting these equations into Eq.~(\ref{signalSSB}), we obtain 
\begin{align}
W(t) &= \exp\left( -2\sum_j \mu_j^2 \bigl|\tilde{f}_t(\omega_j) \big|^2
+ O\bigl((\mu_j / \omega_j)^3\bigr) \right)
\notag\\
&= \exp \left( -2\int_0^\infty {\rm d}\omega 
J(\omega) \big| \tilde{f}_t(\omega) \big| ^2 
+ O\bigl((\mu_j / \omega_j)^3\bigr) \right).
\notag
\end{align}
Using Eq.~(\ref{S_J_spinspin}), we finally obtain the desired result (\ref{signal_S}).

\end{document}